\documentstyle[preprint,aps]{revtex}
\begin{document}
\draft
\title{Perspective: Atomic Parity Violation and the Nuclear Anapole Moment}
\author{W. C. Haxton}
\address{Institute for Nuclear Theory, Box 351550, 
and Department of Physics, Box 351560,\\
University of Washington, Seattle, Washington 98195}

\date{\today}
\maketitle
Until 40 years ago, physicists had assumed that the fundamental
forces of nature did not distinguish between left and right.
That is, it was believed that the laws of physics in a
mirror-symmetric universe would be the same as in ours. 
Then in 1957, following a suggestion by Lee and Yang \cite{lee},
experimenters discovered that the weak force, which is 
responsible for beta decay in nuclei, violated this conservation
of parity \cite{wu}.  Shortly thereafter,
the Soviet physicists V. G. Vaks and
Ya. B. Zeldovich \cite{zeldovich} independently noted that
particles could then have parity-violating
couplings to the electromagnetic field.  Such ``anapole moments"
would arise, in the more modern language of today's standard model
of electroweak interactions, 
from very small effects associated with exchange
of $W^{\pm}$ or $Z^0$ bosons between, for example, the quarks within a nucleon or nucleus.
In this issue of Science, the first definitive
measurement of an anapole moment is reported by the University
of Colorado group of C. S. Wood {\it et al} \cite{wood}.

Some nuclear interactions with the electromagnetic field are
quite familiar.  As a charged object, the nucleus accelerates 
when an electric field is applied.  If the nucleus has a 
nonzero spin $\vec{I}$, it also has an interaction with an
applied magnetic field $\vec{B}$ of the form $\mu \vec{B} \cdot \vec{I}$,
where $\mu$ is the magnetic moment.  More exotic interactions
can arise when symmetries preserved by electromagnetism are
violated by other, weaker forces.  Perhaps the best 
known of these is the electric dipole moment $d_N$, which can be
visualized as an asymmetric distribution of charge along a
particle's spin axis.  A particle with an intrinsic dipole  
moment will experience an interaction $d_N \vec{E} \cdot \vec{I}$ 
when placed in an electric field $\vec{E}$.  Electric dipole moments
arise only if the laws of physics are asymmetric 
under both parity inversion and time reversal; studies of the decays 
of the long-lived neutral K meson have shown that this combination of symmetries is
violated, though only very weakly.  Consequently,
despite considerable effort, no one has succeeded in detecting
a nonzero electric dipole moment.

This had also been the case for the anapole moment, which 
can be generated by parity violation in the weak interaction,
but does not require time reversal violation.
This rank-one moment has a number of curious properties.
It vanishes when probed by real photons, i.e., photons
satisfying the usual energy-momentum relation.  Thus the anapole 
moment of a nucleus, for example,
can be measured only in processes where virtual photons
are exchanged with some interacting particle,
such as an atomic electron.  In can be shown that the resulting
electron-nucleus interaction is point-like: the atomic cloud
feels the nuclear anapole moment only to the extent that
the wave functions of the orbiting electrons penetrate the nucleus.
While the exchanged photon is electric dipole, its absorption
by the nucleus takes place through parity-violating components of
the nuclear wave function.  The combination of the usual nuclear
current and parity violation produces a current configuration
similar to a winding about a torus.  The anapole moment is
associated with the resulting magnetic field induced within
the torus (see figure in Ref. [5]).

Exquisitely precise ($\sim$ 1\%) measurements of atomic parity 
violation have been made in recent years.  It is now widely
recognized that these efforts are important not only as 
tests of the standard electroweak model, determining
parameters such as the weak mixing angle $\theta_W$, but also 
as crucial searches for new physics beyond the standard model, complementing
the efforts at high energy colliders.  The dominant
contribution to atomic parity violation comes from
direct $Z^0$ exchange between the electrons and the 
nucleus, with the electron coupling being axial (or spin-dependent)
and the nuclear coupling vector.  The interaction with the
nucleus is thus coherent, proportional to the total weak charge,
a quantity that scales approximately as the neutron number.

Almost two decades ago it was realized that the electromagnetic
interaction of atomic electrons with the nuclear anapole
moment might generate a measureable nuclear spin dependence
in atomic parity violation experiments \cite{khrip}.
The associated effects, which are considerably
weaker than those of the coherent $Z^0$ interaction,
involve a vector coupling to the atomic electrons and an axial
coupling to the nucleus.  The corresponding parity-violating atomic Hamiltonian is
\begin{equation}
H_W^A = \frac{G_F}{\sqrt{2}} \kappa \vec{\alpha} \cdot \vec{I} \rho (r)
\end{equation}
where $G_F$ is the weak coupling constant, $\kappa$ is a parameter
describing the size of the anapole moment, $\vec{\alpha}$ is the
Dirac matrix operating on the electrons, $\vec{I}$ is the nuclear 
spin, and $\rho(r)$ is the nuclear density, a function of the distance
$r$ from the center of the nucleus.  Such an interaction
can also be generated by direct $Z^0$ exchange similar to that 
described above, but with the electron coupling being vector 
and the nuclear coupling axial vector.  At first it seems very
surprising that the anapole moment could then compete with
this direct contribution: the anapole interaction requires a
photon exchange between electron and nucleus in addition to the
weak interaction within the nucleus.  Such a ``weak radiative 
correction" is naively supressed by a relative factor of the
fine structure constant, 1/137.  However this direct $Z^0$
exchange is inhibited: the axial coupling to the nucleus is
no longer coherent (only the last unpaired nucleon contributes),
and the vector coupling to the electron is quite supressed
due to the factor (4 sin$^2\theta_W -1)/2 \sim -0.05$.
Furthermore, the nuclear anapole moment
has the remarkable property that it grows as $A^{2/3}$, where
$A$ is the atomic number, thus increasing in proportion to
the nuclear surface area.  The net result is the expectation 
that this weak radiative ``correction" will exceed the direct
nuclear spin-dependent parity violation for nuclei heavier 
than $A \sim 20$.

The problem of separating the anapole moment contribution
from the much larger coherent $Z^0$ exchange remains.
As the former is nuclear spin-dependent while the latter is spin
independent, in principle this separation can be done by 
studying the dependence of the parity violation signal on the
choice of hyperfine level.  In practice, the hyperfine differences are 
very small and their extraction requires heroic efforts to 
control experimental systematics.  An earlier effort by the
Colorado group \cite{Noecker}, where parity violation in $^{133}$Cs
was measured to 2.2\%, provided a tentative identification of
the anapole moment, while the Seattle group's \cite{vetter} 1.2\% measurement
in $^{205}$Tl found a null result, despite reaching a sensitivity
where theorists had predicted an effect.  The unprecedented precision of the
measurements reported in this issue by Wood {\it et al}. \cite{wood}, a seven-fold 
improvement to 0.35\% in the $^{133}$Cs results,
has produced the first definitive isolation of nuclear-spin-dependent
atomic parity violation.  The resulting value for $\kappa$ in
Eq. (1) is 0.127 $\pm$ 0.019, a result differing from zero by $\sim$ 7$\sigma$.  (The atomic
matrix elements of Ref. \cite{blundell} were used in this determination.)
  
One can write $\kappa = \kappa(Z^0) \left[ 1 + {\kappa(A) \over \kappa(Z^0)} \right]$,
where $\kappa(Z^0)$ is the nuclear-spin-dependent $Z^0$ exchange
contribution and the quantity is parenthesis is the expected
enhancement due to the anapole weak radiative correction.  The
standard model gives
\begin{equation}
\kappa(Z^0) = - {G_A \over 2} (1-4 \mathrm{sin}^2\theta_W) {\langle I \| \sum_{i=1}^A \sigma(i) \tau(i) \|I\rangle
\over \sqrt{(2I+1)(I+1)I}} \sim 0.0132
\end{equation}
where $G_A \sim$ 1.26 and sin$^2\theta_W \sim$ 0.223.  The nuclear 
matrix element is taken from the shell model calculation of
\cite{haxton}, which is in good agreement with the single-particle estimate
of \cite{flambaum}.  Clearly the $^{133}$Cs result demands an
additional source of spin-dependent atomic parity violation,
the anapole moment.

As the calculations of \cite{haxton,flambaum} show that the largest
contribution to the nuclear anapole moment arises from parity
mixing in the nuclear wave function,
the efforts of Wood {\it et al} have produced a new technique for
studying the hadronic weak interaction.  This interaction has
proven more elusive than the weak interactions involving
leptons.  While the charged current hadronic weak interactions can be
studied in strangeness- or charm-changing decays, the
standard model predicts that neutral current interactions do
not change flavor.  Thus the only opportunity for studying
the hadronic interactions of the $Z^0$ is provided by nucleon-nucleon
interactions, where parity violation must be exploited to
separate the weak interaction from the much stronger strong
and electromagnetic interactions.
But this is a tough game: only a few experiments 
have been done with the precision required to see an effect,
and only some of the nuclear systems
are sufficiently well understood to allow a quantitative interpretation 
of the results \cite{adelberger}.
Thus new atomic physics techniques, applicable to a variety of
nuclei, could have substantial impact on this field.

When the effects of the $^{133}$Cs anapole moment are included,
the calculations of \cite{haxton,flambaum} yield values
of $\kappa$ of 0.074 and 0.074-0.095, respectively.  At a 
qualitative level, these results are quite pleasing:
theory predicts that radiative corrections will 
strongly enhance nuclear-spin-dependent atomic parity violation,
and the magnitude of the predicted enhancement is in reasonable accord
with the measurements of Wood {\it et al}.  
However these calculations employed ``best value" hadronic weak meson-nucleon
couplings of \cite{donoghue}, while
experimental evidence has mounted \cite{adelberger} that these
couplings may be somewhat too high.  The $^{133}$Cs anapole moment 
depends \cite{haxton} primarily on the coupling combination 
$(f_{\pi} + 0.52f_{\rho}^0)$, where $f_{\pi}$ and $f_{\rho}^0$
are the pion and isoscalar $\rho$ weak meson-nucleon couplings in units of 
the best values \cite{donoghue}.  Hadronic parity violation experiments suggest that $f_{\pi}$
may be substantially less than one, a puzzling and important result as this
coupling is generated almost entirely by the neutral current.
But the $^{133}$Cs anapole results do not appear
consistent with this conclusion.  This is a conflict that
will clearly draw some attention.
  
\pagebreak

\pagebreak

\end{document}